# Quantized Auger Recombination of Polaronic Self-trapped Excitons in Bulk Iron Oxide

Hongyan Liao#,[1] Yunyan Fan#,[1] Yumei Lin,[1] Kang Wang,[1] Renfu Li,[2] Xueyuan Chen,[2]

Kelvin H. L. Zhang*,[1] Ye Yang*[1]

1. State Key Laboratory of Physical Chemistry of Solid Surfaces, College of Chemistry and Chemical Engineering, Xiamen University, Xiamen 361005, China
2. CAS Key Laboratory of Design and Assembly of Functional Nanostructures and Fujian Key Laboratory of Nanomaterials, Fujian Institute of Research on the Structure of Matter, Chinese Academy of Sciences, Fuzhou 350002, China

#These authors contributed equally.
*To whom correspondence should be addressed.
kelvinzhang@xmu.edu.cn; ye.yang@xmu.edu.cn

**Abstract:** The Auger recombination in bulk semiconductors can depopulate the charge carriers in a non-radiative way, which, fortunately, only has detrimental impact on optoelectronic device performance under the condition of high carrier density because the restriction arising from concurrent momentum and energy conservation limits the Auger rate. Here, we surprisingly found that the Auger recombination in bulk α-Fe$_2$O$_3$ films was more efficient than narrow-bandgap high-mobility semiconductors that were supposed to have much higher Auger rate constants than metal oxides. The Auger process in α-Fe$_2$O$_3$ was ascribed to the Coulombically coupled self-trapped excitons (STEs), which was enhanced by the relaxation of momentum conservation because of the strong spatial localization of these STEs. Furthermore, due to this localization effect the kinetic traces of the STE annihilation for different STE densities exhibited characteristics of quantized Auger recombination, and we demonstrated that these traces could be simultaneously modeled by taking into account the quantized Auger rates.

Metal oxides offer opportunities for stable and cost-effective solar-to-fuel conversion *via* water splitting or artificial photosynthesis. Because of the strong electron-phonon interaction that arises from the ionic and polar nature of the metal oxides, a charge carrier can displace equilibrium positions of the surrounding ions so as to create a potential well that traps the carrier itself.(*1-4*) This self-trapping phenomenon has been suggested to be responsible for the insufficient characteristics of metal oxides, such as low mobility,(*5, 6*) short lifetime(*7-9*) and chemical potential drop of photocarriers(*10, 11*). Therefore, fundamental understanding of the self-trapped carrier dynamics regarding the formation, transport and recombination is of key importance in the development of the metal oxide based solar-to-fuel generation, which is attracting a great deal of attention.(*5-7, 9, 12-16*)

A self-trapped electron (hole) together with the accompanying lattice deformation is often regarded as a quasi-particle, an electron (hole) polaron. Generally, polarons in different media are categorized into two classes, large polarons and small polarons. The former spatially extends over many lattice cells, while the latter is confined within an atomic dimension. These two types of polarons manifest distinct photophysical properties. Despite its large effective mass, the large polaron possesses a sizable mobility (several to tens $cm^2V^{-1}s^{-1}$) because it is weakly scattered by phonons. Thus, the large polaron has been suggested to account for the slow hot carrier cooling and the fairly large mobility in lead halide perovskites.(*17, 18*) In contrast, the mobility of a small polaron is usually very low because the transport is dominated by a succession of the phonon assisted hops between adjacent self-trapping sites.(*14, 19*) In metal oxides, free carriers tend to form small polarons through quick self-trapping,(*3, 4, 13, 15*) and recent ultrafast spectroscopic studies have confirmed that this transient transition occurs on a sub-ps time scale.(*7, 9, 16, 20, 21*)

Furthermore, due to the Coulomb attraction, the photogenerated electron and hole in metal oxides could form an energetically favorable electron-hole pair confined within a self-trapping potential well, which is usually referred to as a polaronic self-trapped exciton (STE).(*2*) Highly localized STEs in several metal oxides have already been experimentally detected.(*12, 22*) Because long-range interactions between these neutral quasi-particles are negligible, dynamic behaviors of STEs are expected to be distinct from those of mobile excitons. Although the STE dynamics determines photocarriers' transport, lifetime and other important parameters relevant to the energy conversion efficiency, the STE annihilation mechanism, especially that involves multiple STEs, in bulk metal oxides still remain unknown.

Here, a single crystalline thin film of $\alpha$-Fe$_2$O$_3$ was employed as a platform to study the annihilation mechanism of highly localized STEs. The dynamics of STEs was measured by probing the kinetics of transient absorption (TA) features recorded in the region well-below the bandgap of $\alpha$-Fe$_2$O$_3$. The obtained STE annihilation curves for different STE densities exhibited a biphasic decay fashion, a fast component sensitive to the STE density and a common slow component independent to the STE density. These kinetics traces could not be modeled by the conventional charge recombination mechanisms. Inspired by the multiexciton recombination model for quantum dots, we attributed the fast STE annihilation to the quantized Auger recombination of the electronically coupled STEs. We demonstrated that by including discrete Auger rates, STE annihilation kinetics for different STE densities could be simultaneously simulated with only two fitting parameters.

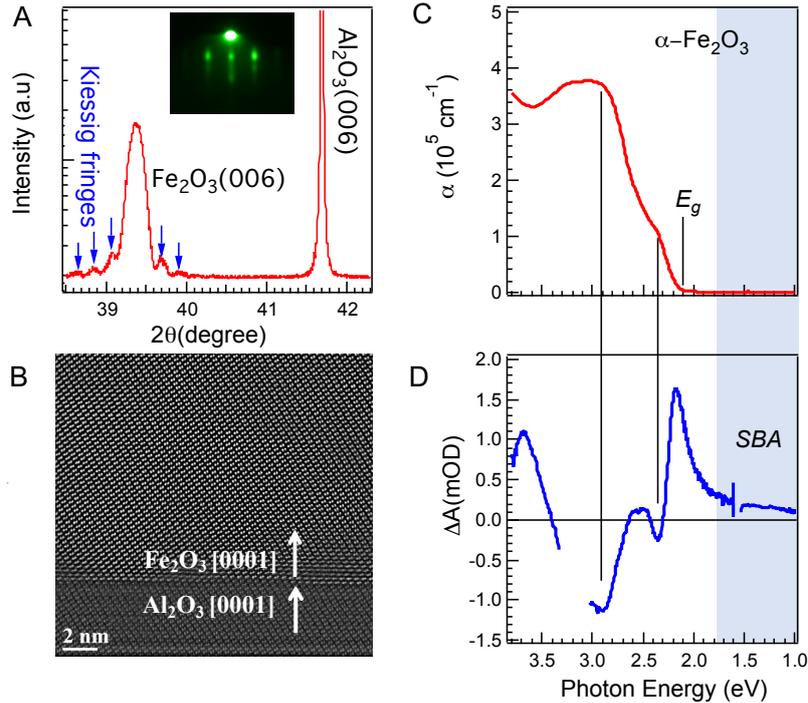

**Figure.1 Structural and spectral characterizations of the α-Fe$_2$O$_3$ film.** (A) XRD pattern of the α-Fe$_2$O$_3$ thin film grown on an Al$_2$O$_3$ (0001) substrate. Kiessig fringes are indicated by the blue arrows. The streaky RHEED pattern (inset) indicates a smooth surface. (B) Cross-section STEM image of the α-Fe$_2$O$_3$ /Al$_2$O$_3$ (0001) interface, viewed along [2$\bar{1}\bar{1}$0] Al$_2$O$_3$ direction. (C) The absorption coefficient as function of photon energy. The bandgap energy ($E_g$) is labeled. (D) A representative transient absorption (TA) spectrum. The region shaded blue indicates the deep sub-bandgap region, and the sub-bandgap absorption (SBA) is associated with electrons. Panel C and D share the same x-axis.

The α-Fe$_2$O$_3$ thin films with thickness of around 40 nm were epitaxially grown on Al$_2$O$_3$ (0001) substrates by pulsed laser deposition, and the growth was monitored by *in-situ* reflection high-energy electron-diffraction (RHEED). The θ-2θ X-ray diffraction (XRD) out-of-plane scan of the sample shows the Fe$_2$O$_3$ (0006) reflection with Kiessig fringes (Fig. 1A), suggesting the high crystalline quality of the epitaxial film with well-defined surface and interface. The Fe$_2$O$_3$ film is atomically flat, evidenced by the RHEED pattern (inset of Fig 1A) and atomic force microscopy (AFM) measurement (Fig. S1A). Scanning transmission electron microscopy (STEM) images (Fig. 1B and Fig. S1B) further justify the atomic uniformity of the thin film. Reciprocal space maps (RSM) near the ($\bar{1}$010) reflection of Fe$_2$O$_3$ and Al$_2$O$_3$ reveals the lattice parameters (a=5.034 Å, c= 13.735 Å) of the film (Fig. S1C), close to those of bulk α-Fe$_2$O$_3$.(*23*) The details of the sample preparation and characterization were described in Supplementary Text 1.

The absorption coefficient of the sample (Fig. 1C) measured by ellipsometry shows an onset at approximately 2.1 eV, corresponding to the bandgap of α-Fe$_2$O$_3$. The two absorption peaks at photon energy ($hv$) of 3.0 eV and 2.3 eV have been attributed to a ligand-to-metal charge transfer band ($O^{2-}\ 2p \rightarrow Fe^{3+}\ t_{2g}^*$) and iron d-d transitions, respectively.(*8*) For comparison, a typical TA spectrum is also plotted in the same spectral range (Fig. 1D). As indicated by the two vertical lines, the photobleaching arises from the carrier occupation at the band edges, which weakens the corresponding absorption peaks due to the Pauli blocking effect.(*8, 24, 25*) Additionally, the redshift of the absorption spectrum due to the bandgap renormalization(*26*) and thermal effect(*8*) should account for the positive peaks and partials of the negative peaks in the TA spectrum. Thus, TA features above the bandgap are caused by the combined effects of Pauli blocking and spectral shifting. Note that only spectral shifting cannot give rise to any signals at the deep sub-bandgap region ($hv$<1.8 eV), which has been justified in the thermo-absorption(*8*) and electro-absorption spectra(*25*). Nonetheless, a broad positive band prevails throughout the deep sub-bandgap region in the TA spectrum (shaded blue, Fig.1D), which has also been observed in previous reports(*8, 9, 27, 28*). This sub-bandgap absorption (SBA) feature coincides with the absorption spectral change caused by electron injection from highly reductive radicals(*29*) or cathodic bias(*25*), implying that the SBA results from the optical transitions associated with photoinjected electrons. In contrast to the complex origin of super-bandgap features, the SBA can be exploited as a convenient probe to measure the photocarrier dynamics.

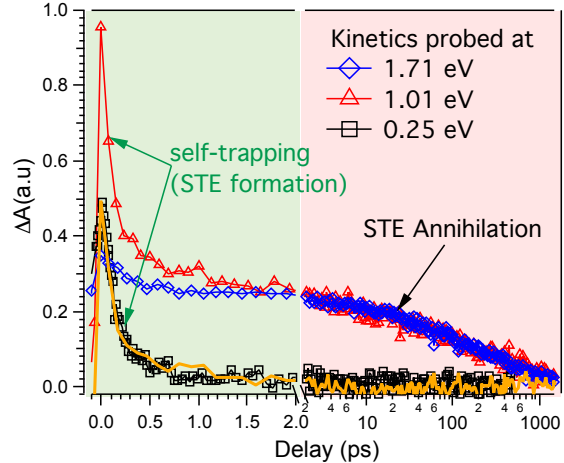

**Figure 2. Formation and annihilation dynamics of self-trapped excitons (STE) in α-Fe$_2$O$_3$.** STE dynamics is represented by the transient absorption (TA) kinetics monitored at different photon energies below the α-Fe$_2$O$_3$ bandgap. The horizontal axis is split into two parts. The first part (-0.5-2 ps, shaded green) is plotted in a linear scale, and the second part (2-1500 ps, shaded red) is plotted in a logarithmic scale. The orange curve represents the difference between the red-triangle and blue-diamond curves. The TA kinetics before and after 2 ps were attributed to the STE formation (free carrier self-trapping) and STE annihilation, respectively.

To explore the nature of the photocarriers in α-Fe$_2$O$_3$, we analyzed the SBA kinetics probed at different sub-bandgap regions under the identical pump condition (Fig. 2). Both of the SBA kinetic traces probed at 1.71 and 1.01 eV show a fast decay completed within 2 ps, which is followed by a relatively slow decay. These slow decay segments in both traces are consistent, while the relative amplitude of the fast component is greater for the kinetics monitored at lower $h\nu$. When $h\nu$ of the probe is tuned downward into the mid-IR region ($h\nu$ =0.25 eV), the obtained kinetics only consists of the fast decay component (black square), which matches the difference of the two SBA kinetics (orange curve). The mid-IR TA probes the intraband transitions of free carriers,(*25*) and thus we attribute the fast component to free carrier self-trapping. The intraband absorption coefficient of free carriers decreases as the $h\nu$ increases, which elucidates the $h\nu$-varying trend of the fast component amplitude. The SBA feature after 2 ps is then ascribed to the absorption of Fe(II) as a consequence of the electron self-trapping at Fe (III) sites, i.e., formation of small electron polarons.(*8, 29*) The time constant of electron self-trapping is then determined to be 0.17±0.01 ps from the single exponential fitting of the mid-IR kinetics (Fig. S2), in line with the time constant of the electron polaron formation measured by the transient XUV spectroscopy.(*7, 11, 20*) Contrary to the electrons, theoretical calculation suggested that hole polarons in α-$Fe_2O_3$ were energetically unfavorable.(*4*) However, owing to the Coulomb attraction, the holes can bind to the small electron polarons to form STEs. It should be noticed that, in addition to the direct Coulomb attraction, the electron polaron can also interact with the hole indirectly through lattice vibrations.(*2*) Highly localized excitons in α-$Fe_2O_3$ have also been experimentally confirmed.(*12*) Thus, the gradual decay of SBA after 2 ps is assigned to the STE annihilation *via* different charge recombination mechanisms that will be discussed in the subsequent sections.

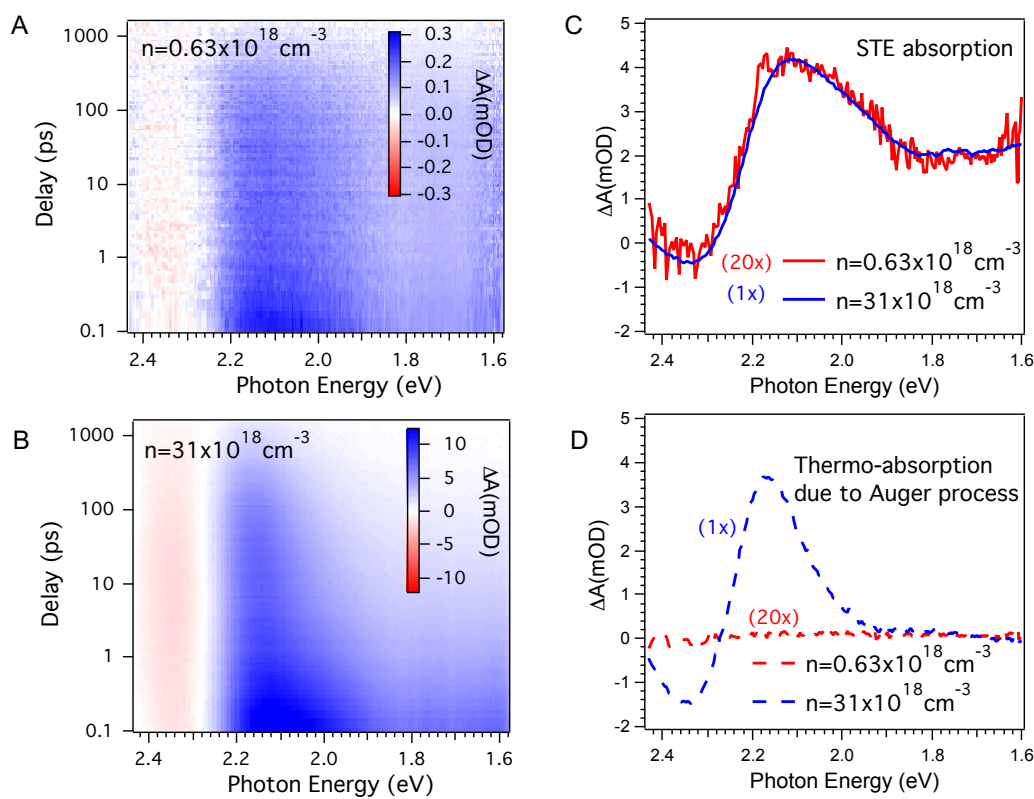

**Figure 3. Dependence of transient absorption (TA) spectra on self-trapped exciton (STE) density.** Pseudo-color images of the TA spectra for low (A) and high (B) pumping intensities, corresponding to STE density of 0.63 and $31 \times 10^{18} cm^{-3}$, respectively. The vertical and horizontal axes represent the pump-probe delay and probe photon energy, respectively. The TA signal magnitude is reflected by the color intensity as indicated by the color scale bar. Spectral components of STE absorption (C) and thermo-absorption (D) derived from TA spectral decomposition. The magnitude of spectral components for the low STE density was magnified by a factor of 20 for better comparison.

The SBA spectral evolution and kinetics for different STE densities were examined in order to interrogate the STE annihilation mechanisms. The STE density was controlled by adjusting the incident pump intensity (see Supplementary Text 1 for details). The pseudo-color images of TA spectra for the lowest and highest STE densities are displayed in Fig. 3, and those for the rest STE densities are shown in Fig. S3. In the case of the lowest STE density ($0.63 \times 10^{18}$ $cm^{-3}$), the TA spectral shape persists when its magnitude declines (Fig.3A), characteristic of a homogenous decay behavior at different probing regions. On the contrary, for the highest STE density ($31 \times 10^{18}$ $cm^{-3}$), the magnitude decay is accompanied by spectral narrowing (Fig. 3B), implying that the TA signal closer to bandgap decays slower than that further from bandgap. This heterogeneous spectral evolution behavior is analyzed through spectral decomposition. Besides the spectral component that mirrors to the TA spectrum for lowest STE density (Fig. 3C), we also find an additional component (Fig. 3D) that resembles the thermo-absorption spectrum,(*8*) which is tentatively attributed to lattice heating caused by non-radiative Auger recombination at high STE density. Because the thermo-absorption component does not comprise any SBA signals, the SBA kinetics is not affected by the spectral narrowing. Additionally, the plot of the SBA magnitude recorded immediately after STE formation (t=2ps) displays a linear relationship with the initial STE density (Fig. S4). Thus, within the range of STE density in current study, the SBA kinetics is always valid to represent the STE dynamics in proportion.

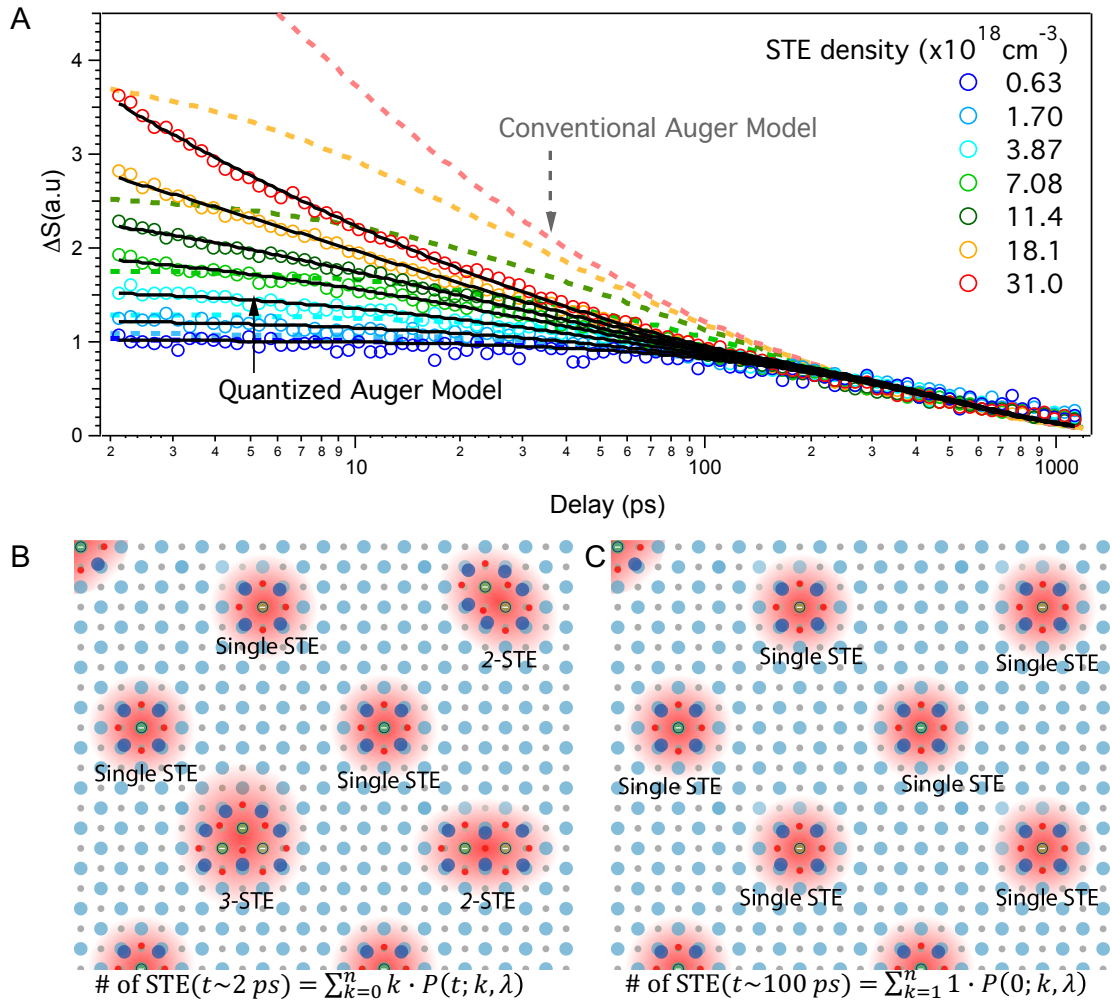

**Figure 4. Density dependent self-trapped exciton (STE) dynamics and illustration of quantized Auger model.** (A) Normalized STE annihilation kinetics for the corresponding STE densities. The colored dashes and black curves correspond to the fitting traces based on the conventional and quantized Auger models, respectively. The delay window ($t<2$ ps) showing the self-trapping process is truncated in order to avoid confusion. Schematic illustration of the STE distribution before (B) and after (C) multi-STE Auger recombination. The small circles labelled by minus signs together with the surrounding lattice distortion represent the small electron polarons, while the areas shaded red represent the bound holes. The Auger recombination takes place within multi-STEs rather than among isolated STEs. The total quantity of the STEs in the two scenarios are formulated as function of the Poisson probabilities (equations underneath the panels). The quantity of the remaining STEs is equal to the quantity of the self-trapping sites. The single STE annihilation within 100 ps due to geminate recombination is not indicated in the illustrations, but it has been taken into account in the data analysis.

The STE annihilation dynamics for different STE densities are represented by the corresponding SBA kinetics after 2 ps, which all exhibit a biphasic decay fashion, a fast component finished within 100 ps followed by a slow component lasting longer than 1 ns. The decay rate of the fast component increases as the STE density rises, while the decay rate of the slow component is independent of the STE density. To emphasize the dependence of the fast component on the STE density, these kinetic traces were scaled so that the amplitude of their slow components were normalized to unity (Fig. 4). The original kinetics and the scaling factors are provided in Fig. S5.

Higher order recombination channels, such as bi-molecular and/or Auger recombination, usually introduce faster decay component under higher intensity excitation. However, these recombination mechanisms are associated with the itinerant particles, such as mobile excitons or free carriers. On the other hand, previous study has demonstrated that polarons in α-Fe$_2$O$_3$ have to take several ns to travel a distance of only few nm,(*5*) so the STEs are nearly static on a time scale of tens ps. Thus, these conventional recombination mechanisms should be invalid for STEs, and as expected, the attempt to model the STE dynamics relying on these mechanisms is not successful. First, the conventional model fails to predict the decay trends of the kinetics (Fig. 4A, dashed curves and Fig. S6), which underestimates (overestimates) the Auger contribution at low (high) STE density. Second, this model also fails to give a reasonable Auger rate constant. Despite the bad simulation, it is worth noting that, to minimize the fitting deviation, the most approximated fitting result gives a very large rate constant for Auger recombination ($C\sim 10^{-28} cm^6 \cdot s^{-1}$, Supplementary Text 2). If we further consider the factors that could reduce the Auger rate in α-Fe$_2$O$_3$,(*30*), such as the wider bandgap, larger dielectric constant and carrier self-trapping, the Auger rate constant for STEs should be smaller than that in the high-mobility semiconductors (e.g., $C\sim 10^{-31} - 10^{-30} cm^6 \cdot s^{-1}$ for Si,(*31*) GaAs,(*32*) InP,(*32*) etc.). From this perspective, the model has to adopt an Auger rate constant exaggerated by orders of magnitude in order to approach the measured kinetic traces. In other words, the Auger rate in α-Fe$_2$O$_3$ appears enhanced compared with the rational deduction.

On the other hand, if the STEs are close enough that their wavefunctions overlap, then the Auger recombination mediated by Coulomb interaction among these coupled STEs can also result in fast STE annihilation. Nevertheless, the Bohr radius ($a_{STE}$) of the STE in α-Fe$_2$O$_3$ was estimated as small as 0.26 nm based on the STE Mott transition density.(*12*) According to the average STE densities, the average distances ($d$) between two STEs were estimated to be in a range of 6-23 nm, an order of magnitude larger than $a_{STE}$. Thus, STEs should be heterogeneously rather than uniformly dispersed in the lattice space (Fig.4B), otherwise the large STE interspace would prevent STEs from electronic coupling. Just as the electrons are shared by the two hydrogen atoms after forming a hydrogen molecule, the holes should be delocalized and shared by the localized electron polarons within a complex of the coupled STEs. An isolated complex of the coupled STEs is then denoted as a multi-STE. More specifically, a multi-STE consisting of $k$ ($k$=2, 3, 4…) coupled STEs is referred to as a $k$-STE. The lattice site, where either a single STE or a multi-STE resides, is defined as a self-trapping site, so the quantity of self-trapping sites is equal to the total quantities of multi-STEs and single STEs (Fig.4B and 4C).

This coupled STE model can explain the STE density dependent kinetics. The convergence of the kinetic traces in Fig. 4A is the sign of the completion of the fast Auger process in different multi-STEs, and the following consistent segment reflects the slow geminate electron-hole recombination of the remaining single STEs, including initially formed single STEs as well as the ones derived from Auger recombination of the multi-STEs. The amplitude of SBA kinetics is proportional to the total quantity of STEs in single STEs and multi-STEs, while the amplitude of the slow component is only determined by the quantity of those remaining single STEs that is also equal to the quantity of the self-trapping sites. Thus, according to the normalizing procedure, the scaled kinetic traces in Fig. 4A actually represents the ratio of total STEs to trapping sites (or average number of STEs per trapping site). For the lowest STE density, the trapping sites are dominated by single STEs, and thus the initial amplitude of the corresponding kinetics is equal to 1. The single STE lifetime ($\tau_1$) is then determined to be $557\pm 22$ ps from a single exponential fitting of this kinetics (Fig. S7). For higher STE densities, the STE number per self-trapping site always exceeds unity before multi-STEs vanish, and decay pattern of the fast component depends on both the quantities of multi-STEs and their respective lifetimes.

To quantitatively describe the STE annihilation dynamics, we need to uncover the distribution of multi-STEs and single STEs. Immediately after optical excitation, the wavefunctions of free photocarriers are widely spread in the lattice space, and then they quickly collapse into self-trapping potential wells with the dimension of a unit cell. Neglecting the interaction between well delocalized free carriers and highly localized STEs, every self-trapping can be regarded as an independent event, and thus the number of STEs falling into a specified interval of the lattice space should obey Poisson distribution before the Auger recombination takes place. The probability for an interval holding $k$ STEs is given by,

$$P(k,\lambda) = \frac{\lambda^k e^{-\lambda}}{k!} \quad (1)$$

where $\lambda$ is the average number of STEs per interval, which is proportional to the ratio of the STE density to interval volume. It should be noted that the volume of the interval is artificially defined, so for each STE density there always exists some values of the interval volume such that no interval can host more than one trapping sites. With the knowledge of STE distribution, the normalized kinetics in Fig.4, which also represent the average number of STEs per self-trapping site, can then be explicitly formulated as,

$$\Delta S(t;\lambda) = \sum_{k=0}^{n} k \cdot P(t;k,\lambda) \Big/ \sum_{k=1}^{n} 1 \cdot P(0;k,\lambda) \quad (2)$$

where the numerator and denominator stand for the total quantity of STEs (including single and multi-STEs) and the quantity of trapping sites, respectively.

For a multi-STE with finite number of coupled STEs, a cascade of Auger recombination can be assumed, in which a $k$-STE decays sequentially to a $(k-1)$-STE, then to a $(k-2)$-STE, and so on until to the single STE, and these recombination steps correspond to a series of discrete Auger rate constants. The same assumption has also been made to interpret the multi-exciton dynamics in quantum dot ensembles.(*33-35*) The multi-STE annihilation rate is implicitly determined by a set of coupled differential equations, and the general expression of the differential equation is given by,

$$\frac{dP(t;k,\lambda)}{dt} = \frac{P(t;k+1,\lambda)}{\tau_{k+1}} - \frac{P(t;k,\lambda)}{\tau_k} \quad (3)$$

where $\tau_k$ is the expected lifetime of a $k$-STE. The first and second terms at the right side of *Eq.* 3 represent the increment of $k$-STE from the decay of $(k+1)$-STE and decrement of $k$-STE due to its decay, respectively. Two-particle Auger recombination is excluded for a multi-STE because it proceeds between two itinerant excitons with large exciton binding energy.(*35-37*) A multi-STE with three-dimensional confinement, like multi-excitons in quantum dots, should decay *via* three-particle Auger recombination, and thus $1/\tau_k$ can be expressed as,(*38*)

$$\frac{1}{\tau_k} = \frac{1}{4\tau_2} k^2(k-1) \quad (4)$$

where $\tau_2$ is the expected lifetime of the 2-STE. Combining *Eq.* 1-4, we find that $\lambda$ and $\tau_2$ are the only two unknown parameters, and by setting them as fitting parameters, all the normalized kinetic traces can be simultaneously simulated (black curves, Fig. 4A). The fitting parameters and code of the fitting program are provided in Supplementary Text 3.

The comparison of the two fitting models in Fig. 4A suggests that the Auger contribution is greater than what the conventional model predicts at low densities because the Auger recombination is enhanced due to the formation of multi-STEs. Like the Auger recombination in nanocrystals,(*39, 40*) the Auger process in multi-STEs could be facilitated by the relaxation of the momentum conservation due to the strong spatial localization, which might explain the enhanced Auger rate.

The best fitting based on the quantized Auger model gives the 2-STE lifetime ($\tau_2$) as 35±1 ps, which falls within the range of the reported biexciton lifetimes in quantum dots.(*33*) However, this is not an ideal comparison because the Bohr radius of the multi-STE is much smaller than the radii of those nanocrystals. Unfortunately, we could not find the biexciton lifetime in nanocrystals with sub-nm radius. Simple extrapolation of the size dependent biexciton lifetime curve in quantum dots indicates that biexciton annihilation in quantum dots with sub-nm size should be much faster than the 2-STE annihilation in $\alpha$-$Fe_2O_3$.(*33*) In quantum dots, electron and hole wavefunctions overlap efficiently, while in a multi-STE, as the electron polarons are of atomic dimension, the wavefunction overlap between the polarons and bound holes are ineffective, which could suppress the Auger process.(*30, 41*)

The quantization of the Auger process in bulk $\alpha$-$Fe_2O_3$ stems from the strong spatial localization of the coupled STEs due to the presence of small polarons. As small polarons have also been reported in a wide variety of metal oxides,(*1, 3, 4, 13, 16, 21*) this quantized Auger recombination should be potentially a general phenomenon in these materials.

**Acknowledgements.**

X.C. acknowledges the support from the Strategic Priority Research Program of the Chinese Academy of Sciences (XDB20000000) and the CAS/SAFEA International Partnership Program for Creative Research Teams. K.H.L. Zhang is supported by National Natural Science Foundation of China under Grant No. 21872116. Y.Y is supported by National Natural Science Foundation of China under Grant No. 21973078.

Supplementary Materials

# Quantized Auger Recombination of Polaronic Self-trapped Excitons in Bulk Iron Oxide

Hongyan Liao#,[1] Yunyan Fan#,[1] Yumei Lin,[1] Kang Wang,[1] Renfu Li,[2] Xueyuan Chen,[2]

Kelvin H. L. Zhang*,[1] Ye Yang*[1]

1. State Key Laboratory of Physical Chemistry of Solid Surfaces, College of Chemistry and Chemical Engineering, Xiamen University, Xiamen 361005, China
2. CAS Key Laboratory of Design and Assembly of Functional Nanostructures and Fujian Key Laboratory of Nanomaterials, Fujian Institute of Research on the Structure of Matter, Chinese Academy of Sciences, Fuzhou 350002, China

#These authors contributed equally.
*To whom correspondence should be addressed.
kelvinzhang@xmu.edu.cn; ye.yang@xmu.edu.cn

**Supplementary Figures.**

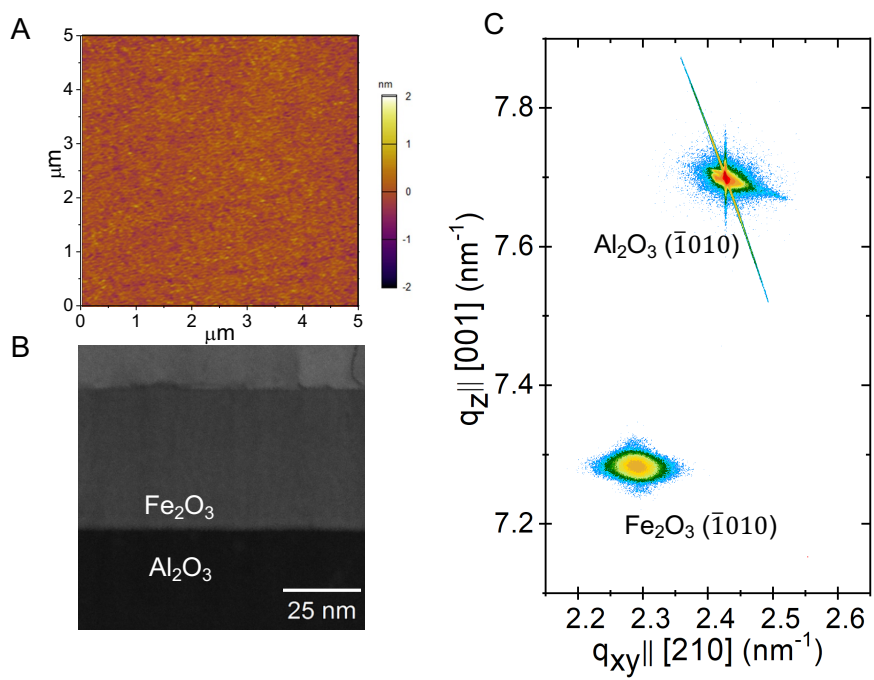

**Figure S1. Structural characterization for the α-Fe$_2$O$_3$ thin film.** (A) Atomic force microscopy (AFM) image of the Fe$_2$O$_3$ film with a root mean roughness of 0.2 nm. (B) Large area cross-sectional STEM image of the Fe$_2$O$_3$/Al$_2$O$_3$ (0001) interfaces. (C) A typical reciprocal space map (RSM) of ~40 nm thick α-Fe$_2$O$_3$ epilayer films around the ($\bar{1}$010) reflections, with longitudinal transfer along [001] in the vertical direction and transverse wavevector transfer along [210] direction.

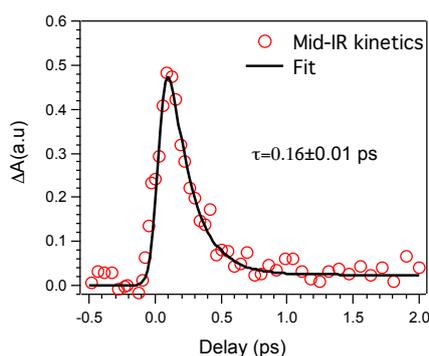

**Figure S2. Single exponential fitting of the mid-IR kinetics.** The fitting curve is the convolution of the instrument response function (a Gaussian function) and a single exponential decay function. The full width at half maximum (FWHM) of the Gaussian function reflects the instrument response time, which is 130 fs for the current transient absorption setup.

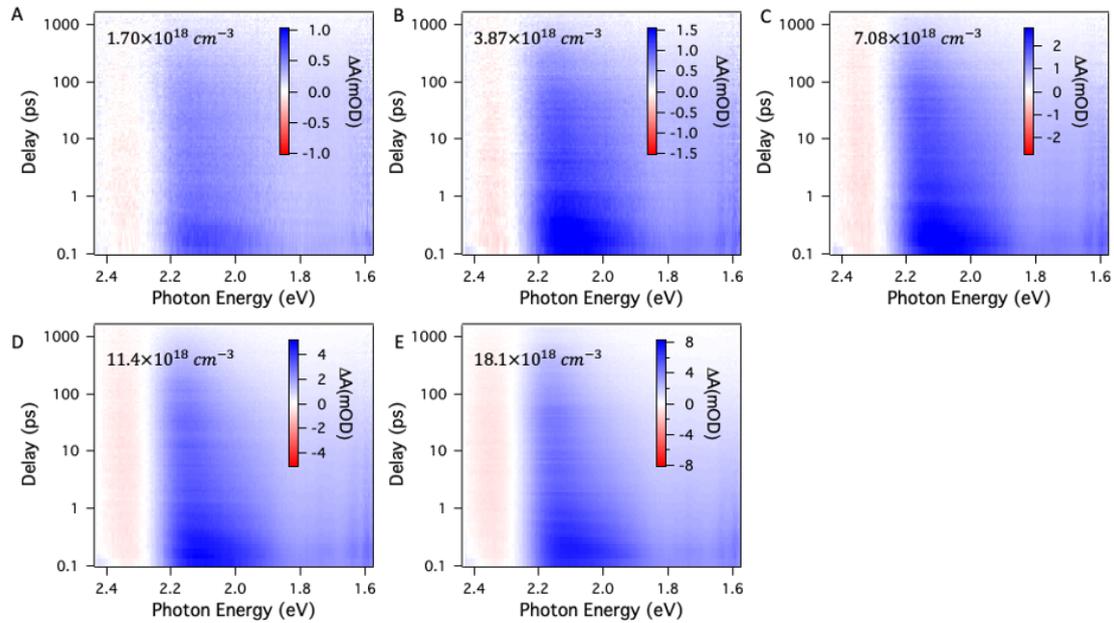

**Figure S3. Display of the pseudo-color images of the transient absorption spectra for different STE densities.** Those for lowest and the highest STE densities are shown in the main text. The vertical and horizontal axes represent the pump-probe delay and probe photon energy, respectively. The TA signal magnitude is reflected by the color intensity as indicated by the color scale bar. The corresponding STE density is indicated in each panel.

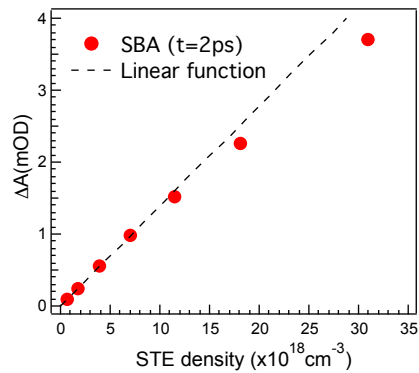

**Figure S4. The linear relationship between sub-bandgap absorption (SBA) and STE density.** The black solid line is a linear function with zero intercept. The SBA data was recorded after the completion of self-trapping but before the annihilation of the STEs (at delay of 2ps). The small deviation between experimental data and the linear function at high STE densities is probably due to slight multi-STE annihilation before 2 ps.

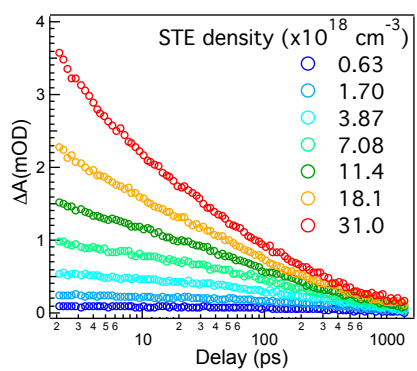

**Figure S5. Original SBA kinetics for different STE density.** The normalized kinetic traces shown in Fig.4 in the main text are obtained from linear scaling of the SBA kinetics in Fig. S4. The corresponding scaling factors are $1.13 \times 10^4$, $5.09 \times 10^3$, $2.87 \times 10^3$, $1.96 \times 10^3$, $1.50 \times 10^3$, $1.24 \times 10^3$ and $1.02 \times 10^3$ for the indicated STE densities from the lowest to the highest.

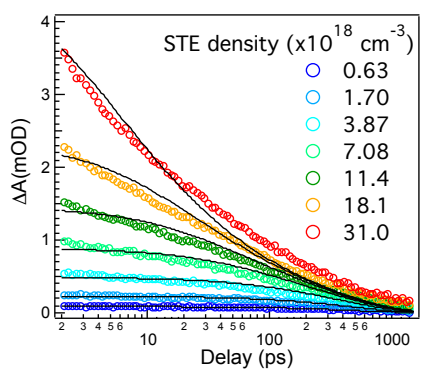

**Figure S6. Fitting of the density dependent STE annihilation kinetics based on the conventional carrier recombination model.** The fitting procedure is detailed in the Supplementary Text 2. The comparison with most approximated fitting curves indicates that the model overestimates (underestimates) contribution from fast (slow) component for the higher STE densities (11.4, 18.1 and 31.0 $\times 10^{18}\ cm^{-3}$), but the deviations between the data and model was inversed for the lower STE densities (0.63, 1.70 and 3.87 $\times 10^{18}\ cm^{-3}$).

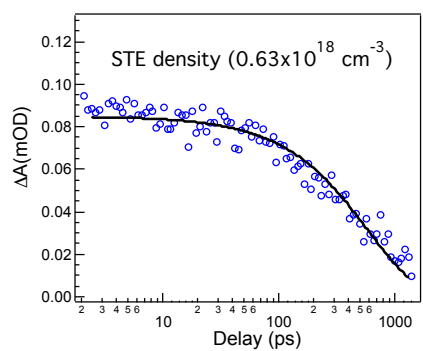

**Figure S7. Single exponential fitting of the SBA kinetics for the lowest STE density.** Because the multi-STE is assumed to be negligible small in this case, the SBA kinetics represents the geminate recombination (a type of monomolecular recombination

mechanisms) of the single STEs. The lifetime of the single STEs is then determined to be 557± 22 ps from this fitting.

**Supplementary Text.**

**Text 1. Sample preparation and characterization methods.**

**Growth of α-Fe$_2$O$_3$ films.** The α-Fe$_2$O$_3$ epitaxial thin films with thicknesses of 40 nm were grown on Al$_2$O$_3$ (0001) substrates by pulsed laser deposition (PLD) method, and the film growth was monitored by *in-situ* reflection high-energy electron-diffraction (RHEED). The α-Fe$_2$O$_3$ target was prepared by grinding and pressing high-purity Fe$_2$O$_3$ (99.99%, Alfa Aesar) and sintered in air at 1250°C for 12 hours. The laser ablation was carried out using a 248 nm KrF excimer laser with an energy density of 1.2 J/cm$_2$ and a pulse repetition rate of 5 Hz. The films were deposited at substrate temperature of 550°C in 100 mTorr O$_2$ partial pressure and cooled to room temperature under the same O$_2$ pressure. The crystal structure and epitaxial relationship were determined by high-resolution XRD using a PANalytical four-circle diffractometer in θ-2θ scans and reciprocal space mapping (RSM) modes. The surface morphologies were characterized by AFM in tapping mode. Cross-sectional STEM specimens were prepared with a FEI Helios dual-beam focused ion beam/scanning electron microscope using a standard lift-out approach. A FEI Titan transmission electron microscopy with a spherical aberration corrector for the probe-forming lens operating at 300 kV was used for high-resolution high-angle annular dark-field STEM imaging.

**Transient absorption measurement.** The transient absorption (TA) setup was built based on a regeneratively amplified Ti:sapphire laser system. The center wavelength of the output beam was around 800nm, and the pulse repetition rate of this beam was 1 KHz. The visible probe was generated by focusing an attenuated 800nm beam into a CaF$_2$ window for probing range of 350-650nm or a sapphire window for probing range of 450-780 nm. The mid-IR probe was generated by a TOPAS optical parametric amplifier with a difference frequency generation (DFG) extension. Another 800nm beam with pulse energy of ~2 mJ was sent into a TOPAS optical parametric amplifier to generate pump pulse. The pump beam was chopped at frequency of 500 Hz. The probe pulses were delayed in time with respect to the pump pulses using a motorized translation stage. The pump and probe beam were spatially overlapped on the surface of the sample. The transmitted probe beam was directed to the multichannel CMOS sensors by an optical fiber, and the transmitted pump beam was stopped by a beam blocker. The intensity of the pump beam was tuned by a neutral density filter wheel. The beam size of the probe and pump at the sample position were around 250 μm and 1.5 mm, respectively.

**Determination of STE density.** The incident pump photon flux ($I_0$) was determined by measuring the pump pulse energy after a pinhole with diameter of 400 μm placed at the sample position. The small angle reflectance ($R$) of the sample at 500 nm (pump wavelength) was measured to be 44.5%, and the transmittance ($T$) at the same wavelength was measured to be 29.4%. With assumption that every absorbed photon was converted into an STE, the STE density ($n$) in the α-$Fe_2O_3$ film was then calculated by the following equation,

$$n = I_0 \times (1 - T - R)/L \quad (S1)$$

where $L$ was the film thickness (40 nm in the current study).

**Text 2. Conventional recombination model.** This fitting model is based on the conventional charge recombination equation,

$$\frac{dn}{dt} = -An - Bn^2 - Cn^3 \quad (S2)$$

where $n$ is carrier density, proportional to the kinetics amplitude, and A, B and C are the first (e.g. geminate recombination), second (e.g. bimolecular radiative, non-radiative recombination) and third (e.g. Auger recombination) order recombination rate constants. The fitting program was carried out by Igor Pro software (analysis/Global Fit). The code of the fitting program is provided as followings. As the SBA signal ($\Delta A$) is proportional to the STE density ($n$) with proportionality constant $k$, (i.e., $n = k \cdot \Delta A$), the fitting equation for SBA kinetic traces can be derived from Eq. S2.,

$$\frac{d(\Delta A)}{dt} = -A(\Delta A) - B'(\Delta A)^2 - C'(\Delta A)^3 \quad (S3)$$

where $B'$ is equal to $B \cdot k$, and $C'$ is equal to $C \cdot k^2$. This fitting model was coded in a fitting program that was also carried out by Igor Pro software (analysis/Global Fit). The code of the fitting program (blue text) and interpretations (red text) are provided as followings.

```
#pragma rtGlobals=1		// Use modern global access method.
#include <Global Fit 2>
Function CRM(pw,yw,tt) : fitfunc
 wave pw, yw,tt	//pw is the parameter wave, yw is the kinetics, tt is the delay; pw[0] is A, the first order recombination rate constant; pw[1] is B', the first order recombination
```

rate constant; pw[2] is C', the first order recombination rate constant; pw[3] is the initial amplitude of the kinetics.
  IntegrateODE/X=tt RateEq, pw, yw //The actual fitting curve is the numerical integration of the differential rate equation.

End

Function RateEq(pw,tt,yy,dydt)
  wave pw, yy, dydt
  variable tt // not actually used in this fitting
  variable A = pw[0] // A was obtained independently from a single exponential fitting of the lowest STE kinetics, so this fitting parameter was actually fixed (not varying) during the fitting.
  variable BB = pw[1] // BB is B'
  variable CC = pw[2] // CC is C'
  dydt[0] = -A*yy[0] - BB*yy[0]^2 - CC*yy[0]^3 // Eq. S3
  return 0
End

**Table S1. The list of linked fitting parameters from the global fitting results.**

| Linked parameters | value | fitting uncertainty |
|---|---|---|
| $B'(OD^{-1} \cdot s^{-1})$ * | $6.9 \times 10^{-16}$ | 0.03 |
| $C'(OD^{-2} \cdot s^{-1})$ ** | $7.3 \times 10^{3}$ | $0.1 \times 10^{3}$ |

* The fitting parameter $B'$ is constrained to be positive in order to give a physically meaningful result. As indicated in this table, the resulted value for $B'$ is extremely small with a relatively large fitting uncertainty, and thus we consider that the second order recombination is negligible.
** The Auger rate constant $C$ is then determined to be $1.4 \times 10^{-28} cm^6 \cdot s^{-1}$ from fitting parameter $C'$ and the proportionality constant $k$ ($7.1 \times 10^{21} cm^{-3} \cdot OD^{-1}$).

**Table S2. The list of unlinked fitting parameters from the global fitting results. The**

| STE density ($\times 10^{18} \cdot cm^{-3}$) | unlinked parameter, $pw[3]$ (OD) | fitting uncertainty (OD) |
|---|---|---|
| 0.63 | $0.87 \times 10^{-4}$ | $0.07 \times 10^{-4}$ |
| 1.70 | $2.3 \times 10^{-4}$ | $0.1 \times 10^{-4}$ |
| 3.87 | $4.9 \times 10^{-4}$ | $0.1 \times 10^{-4}$ |
| 7.08 | $8.8 \times 10^{-4}$ | $0.1 \times 10^{-4}$ |
| 11.4 | $14.2 \times 10^{-4}$ | $0.2 \times 10^{-4}$ |
| 18.1 | $22.3 \times 10^{-4}$ | $0.2 \times 10^{-4}$ |
| 31.0 | $39.3 \times 10^{-4}$ | $0.3 \times 10^{-4}$ |

**Text 3. Quantized Auger recombination model.** This fitting model is based on the Eq. 1-4 and is elaborated in the main text. The fitting program was also carried out by Igor

Pro software (analysis/Global Fit). The code of the fitting program (blue text) and interpretations (red text) are provided as followings.

```
#pragma rtGlobals=1        // Use modern global access method.
#include <Global Fit 2>
Function PoissonDynamics(pw,yw,tt) : fitfunc
    wave pw, yw,tt        //pw is the parameter wave; yw is the kinetics; tt is the delay.
```
pw[0] is the first order recombination rate constant to account for the single STE recombination. pw[1] is the 2-STE annihilation rate constant ($1/\tau_2$). pw[2] is $\lambda$ (see main text for the definition).

```
    variable n
    n=dimsize(yw,0)
    Make/D/O/N=((n),10) ChemKin
```
//Generating a matrix to save the Poisson probability for each multi-STE [upto 10-STEs, contributions from k-STE(k>10) is negligible small due to the extremely low distribution probability even for the largest STE density].

```
    chemKin[0][0]=exp(-pw[2])*pw[2]
    chemKin[0][1]=exp(-pw[2])*pw[2]^2/factorial(2)
    chemKin[0][2]=exp(-pw[2])*pw[2]^3/factorial(3)
    chemKin[0][3]=exp(-pw[2])*pw[2]^4/factorial(4)
    chemKin[0][4]=exp(-pw[2])*pw[2]^5/factorial(5)
    chemKin[0][5]=exp(-pw[2])*pw[2]^6/factorial(6)
    chemKin[0][6]=exp(-pw[2])*pw[2]^7/factorial(7)
    chemKin[0][7]=exp(-pw[2])*pw[2]^8/factorial(8)
    chemKin[0][8]=exp(-pw[2])*pw[2]^9/factorial(9)
    chemKin[0][9]=exp(-pw[2])*pw[2]^10/factorial(10)

    IntegrateODE/X=tt RateEq, pw, ChemKin
```
// integration of the coupled differential equation (Eq. 3 in the main text)
```
    variable i
    yw=0
    duplicate/O yw ywtemp
    for (i=0;i<10;i+=1)    //add up the contributions from all k-STE (k≤10)
        matrixOP/O ywtemp=(i+1)*col(ChemKin,i)
        yw=ywtemp+yw
    Endfor

    yw=yw/(1-exp(-pw[2]))
End

Function RateEq(pw,tt,yw,dydt)
    wave pw
    variable tt
    wave yw // here yw[i] is ChemKin[][i]
    wave dydt //integration of dydt[k] will give the contribution from k-STE.
```

```
    variable k1, k2, k3, k4, k5, k6, k7, k8, k9, k10//rate constants for k-STE, which are
function of rate constant of 2-STE (see Eq.4 in the main text). The rate constant of the
single STE (1/τ₁) is determined independently (Fig. S6).
    k1= pw[0]
    k2= pw[1]
    k3= pw[1]/4*3^2*(3-1)
    k4= pw[1]/4*4^2*(4-1)
    k5= pw[1]/4*5^2*(5-1)
    k6= pw[1]/4*6^2*(6-1)
    k7= pw[1]/4*7^2*(7-1)
    k8= pw[1]/4*8^2*(8-1)
    k9= pw[1]/4*9^2*(9-1)
    k10= pw[1]/4*10^2*(10-1)

    dydt[9] =-k10*yw[9]
    dydt[8]=k10*yw[9]-k9*yw[8]
    dydt[7]=k9*yw[8]-k8*yw[7]
    dydt[6]=k8*yw[7]-k7*yw[6]
    dydt[5]=k7*yw[6]-k6*yw[5]
    dydt[4]=k6*yw[5]-k5*yw[4]
    dydt[3]=k5*yw[4]-k4*yw[3]
    dydt[2]=k4*yw[3]-k3*yw[2]
    dydt[1]=k3*yw[2]-k2*yw[1]
    dydt[0]=k2*yw[1]-k1*yw[0]

    return 0
End
```

**Table S3. The list of linked fitting parameters from the global fitting results.**

| Linked parameters | value | fitting uncertainty |
|---|---|---|
| $\tau_2$ (ps) | 35 | 1 |

**Table S4. The list of unlinked fitting parameters from the global fitting results.**

| STE density ($\times 10^{18} \cdot cm^{-3}$) | unlinked parameter, $\lambda$ | fitting uncertainty |
|---|---|---|
| 0.63 | 0.07 | 0.02 |
| 1.70 | 0.40 | 0.02 |
| 3.87 | 0.90 | 0.02 |
| 7.08 | 1.41 | 0.02 |
| 11.4 | 1.89 | 0.02 |
| 18.1 | 2.52 | 0.02 |
| 31.0 | 3.43 | 0.02 |